# Feynman's Path Integral

# and

# Complementarity of One-Quantum Interference and Two-Quanta Interference

GEROLD WALLNER

*Independent researcher at Intellectual Capital and Asset Management GmbH,*
*Wiener Straße 72, 2352 Gumpoldskirchen, Austria*
*mail: gw@gravitino.eu*

August 14th, 2020 (February 14th, 2022)

## ABSTRACT

Quantum information technology is one of the fastest growing sectors within information technology. By misleading interpretations of Bohr's definition of complementarity, the foundation of nonlocal correlations of entanglement is concealed and therefore, such nonlocal effects can apparently not be utilized directly in information technology. As the first step to identify the utilizable nonlocal property of entanglement, this paper adjusts the notion of "complementarity" of first-order and second-order interference. Especially, the nonlocal consequence of the "principle of least action" for entangled systems is shown by the foundation of second-order interference, the shared phase solutions space. Based on the concept of Feynman's path integral, the validity of the theory is demonstrated by applying the theory to well-known experiments, revealing this underlying nonlocal principle. In the next steps of publication, to highlight the relevance of this nonlocal principle for quantum computing and quantum information technology, the "self-consistency principle" of such nonlocal quantum causal correlations will be shown and verified by an experiment.

keywords: interference, entanglement, nonlocal phase state, delayed choice

# I) Introduction

Quantum systems are a wide field of surprise and counter-intuition by the means of non-classical effects. This paper deals with the usage of the misleading concept "first order interference within entangled systems". In predicting the states of a system of entangled quanta, the insights of Feynman's path integral are essential and open a window of understanding "counter-intuitive" EPR paradoxes [1].

Feynman based his interpretation of the path integral of quantum-mechanical amplitudes on the "principle of least action". The key quantity is the action $S$ of the path, e.g., going from $a$ to $b$, which has to be extremal in the dimension energy times time. $L$ is the Lagrangian for the system.

$$S = \int_{t_a}^{t_b} L(\dot{x}, x, t) dt \ . \tag{1}$$

The condition to be an extremum of S is

$$\delta S = S[x + \delta x] - S[x] = 0 \tag{2}$$

to first-order in $\delta x(t)$. In quantum mechanics the amplitude of such a path from $a$ to $b$ is commonly called "kernel" $K(b,a)$. This kernel is the sum over all paths from $a$ to $b$ and results in a phase. Based on the formalism of the Lagrangian, action $S$ is also robust in all kinds of field theories.

$$\phi[x(t)] = K(b, a) = \int_a^b exp\left[\frac{i}{\hbar} S[b, a]\right] Ðx(t) \ . \tag{3}$$

Feynman showed the relation between action $S$ of the path from $a$ to $b$ in units of $\hbar$ and the probability $P(b,a)$ to go from $a$ at time $t_a$ to $b$ at time $t_b$ in form of the absolute square of that amplitude $K(b,a)$. [2]

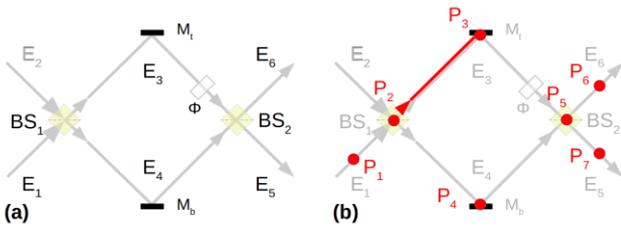

*Fig. 1: A typical MZI is sketched in (a). Different modes are represented by their index. In (b) the path between $P_1$ over $P_3$ to $P_6$ is segmented by points of interest (POI). These POI indicate points of additional phase changing operations like reflections, which has to be considered in the calculations of the path integral.*

To demonstrate these relations a Mach-Zehnder-Interferometer *MZI* with input via mode $E_1$ is introduced (see Fig. 1(a)). For simplicity, the phase of the phase shifter $\Phi$ is assumed to be zero. In the following, this phase shifter is neglected and the label $\Phi$ will be used to demonstrate different effects.

Next, some points of interest (POI) are labeled. A simple example of a path integral within Fig. 1(b) is from point $P_2$ to point $P_3$ in (4).

$$K(P_3, P_2) = \int_{P_2}^{P_3} exp\left[\frac{i}{\hbar} S[P_3, P_2]\right] Ðx(t) \ . \tag{4}$$

To calculate the path integral in a complex arrangement, the following relation defines the composite path integral. Feynman's rule is: "*Amplitudes for events occurring in succession in time multiply.*" [2]

$$S[b, a] = S[b, c] + S[c, a] \ . \tag{5}$$

$$K(b, a) = \int_a^b exp\left[\frac{i}{\hbar} S[b, c] + \frac{i}{\hbar} S[c, a]\right] Ðx(t) \ . \tag{6}$$

A generalized description of aforementioned for the path integral of N points is given by

$$\phi[x(t)] = \prod_{i=1}^N K(i+1, i), \ \forall N = \{2,3,4,\ldots\} \ . \tag{7}$$

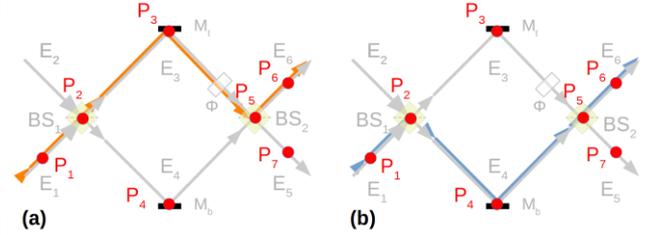

*Fig. 2: (a) Path "top" from $P_1$ to $P_6$ is represented by orange arrows; (b) Path "bottom" from $P_1$ to $P_6$ is represented by blue arrows. Both paths form a spatial superposition. The path integral from $P_1$ to $P_6$ includes both paths, which are the extremal solutions of that specific path integral.*

Reflections at mirrors and reflections at beam splitters contribute additional phases. According to standard photon statistics at beam splitters, the additional and simplified phase is π/2. Including the additional phases of reflections, the amplitude of the path from point $P_1$ to point $P_6$ via point $P_3$ ("top", orange, phase shifter $\Phi$ omitted) can be described by

$$K(P_6, P_1)_{top} = \int_{P_1}^{P_6} exp\left[\frac{i}{\hbar} S[P_6, P_5] + \frac{i*\pi_{BS_2}}{2} + \frac{i}{\hbar} S[P_5, P_3] + \frac{i*\pi_{M_t}}{2} + \frac{i}{\hbar} S[P_3, P_2] + \frac{i*0_{BS_1}}{2} + \frac{i}{\hbar} S[P_2, P_1]\right] Ðx(t) \ , \tag{8}$$

and the path from point $P_1$ to point $P_6$ via point $P_4$ ("bottom", blue) is specified by

$$K(P_6, P_1)_{bottom} = \int_{P_1}^{P_6} exp\left[\frac{i}{\hbar} S[P_6, P_5] + \frac{i*0_{BS_2}}{2} + \frac{i}{\hbar} S[P_5, P_4] + \frac{i*\pi_{M_b}}{2} + \frac{i}{\hbar} S[P_4, P_2] + \frac{i*\pi_{BS_1}}{2} + \frac{i}{\hbar} S[P_2, P_1]\right] Ðx(t) \ , \tag{9}$$

and represented by the "bottom" path.

Either path "top" and path "bottom" are extremal solutions of the path integral $K(P_6,P_1)$. Therefore, both of these paths obey the "principle of least action" and contribute equally to the total complex amplitude. Together, these two paths represent a spatial superposition in a wave equation in (12).

$$\phi_{top}[x(t)] = K(P_6, P_1)_{top} \ . \quad (10)$$

$$\phi_{bottom}[x(t)] = K(P_6, P_1)_{bottom} \ . \quad (11)$$

$$\Psi(P_1, P_6) = \frac{1}{\sqrt{2}}\big(\phi_{top} mode_{top} + \phi_{bottom} mode_{bottom}\big) \ . \quad (12)$$

$$A(P_1, P_6) = \frac{1}{\sqrt{2}}\big(e^{i\phi_{top}} + e^{i\phi_{bottom}}\big) \ . \quad (13)$$

The probability of the path from $P_1$ to $P_6$ is the probability amplitude in (13) squared in (15). Simplifying is done by choosing a new coordinate of $P_6$ by reducing the path length between $P_5$ and $P_6$ in such a way, that $\Phi_{top}$ is zero as it is in (14). This leads to a more convenient form

$$A(P_1, P_6) = \frac{1}{\sqrt{2}}\big(1 + e^{i\phi_{bottom}}\big) \ , \quad (14)$$

and thus,

$$P(P_1, P_6) = \frac{1}{2}\big(1 + cos(\phi_{bottom})\big) \ . \quad (15)$$

If the optical length of these two solutions is equal, only the additional phases of mirrors and beam splitters contribute to the phase difference. Measured after the second beam splitter, interference is part of the formalism of the path integral and depends mainly on these two extremal paths, which are forming a spatial superposition.

**II) Interference and Feynman's path integral**

The nonlocal EPR paradox can be demonstrated by second-order interference in an ex-post analysis. Interference of quanta (e.g., a photon) as a real existing phenomenon can be understood with the theory of the path integral between two points in spacetime. In all cases, in which more than one connecting trajectory between two points are extremal contributors to the path integral (more than one local extremum), we will find a spatial superposition (e.g., double slit or trajectories between laser source and a detector of a Mach-Zehnder-Interferometer MZI).

From the perspective of quantum mechanics even the single solution of the Lagrangian in classical mechanics is a solution of an integral of all paths with their specific phases, which represents interference. So, in quantum mechanics there is always interference, even for just one quantum and a simple transition from point *a* to *b*. Later on in this paper, interference is discussed in the sense of more than one extremal contribution of the path integral of a single quantum or of entangled quanta.

The phase difference of two extremal trajectories, or by Feynman "*interfering alternatives*" [2], can be measured after e.g., the second beam splitter of a MZI, which, of course, contributed already additional phases to these trajectories corresponding to the transformation matrix of this beam splitter. To comply the "principle of least action", the relative phase difference of such two extremal trajectories determines the probability of which out port of the beam splitter has to be taken.

In a double slit experiment, each point in the projection plane is linked with the source by two extremal trajectories or two most contributing parts of the path integral or two local extremal solutions of the "principle of least action". Another term for this is "spatial superposition". Without additional knowledge about the state of the quantum, there is no which-way information. With these prerequisites complied, interference is an observable effect. Interference is just the consequence of solving the path integral including a spatial superposition.

As a spatial superposition is the sum of two extremal paths, the mechanism of the path integral also leads to intensity. By summing up the corresponding phases within each of the two paths, the difference between these resulting phases corresponds with the probability amplitude, including this spatial superposition (0 = maximal intensity; π = minimal intensity). The representation of the extremal trajectories is a helpful reduction of the complexity of the underlying path integral (see Fig. 2).

**III) Interference and entanglement**

Entanglement constitutes an impartible system of quanta. The following discussion focuses on two particle systems, generalizations to multiparticle systems is straight forward. As we learned by Feynman, the rule for a single quantum is: "*Amplitudes for events occurring in succession in time multiply.*" [2]

This rule is expanded to entangled quanta by replacing the mirror of the MZI with a source of entangled quanta and, for further use, adding a second phase shifter $\Phi_1$ into the path *A* or into the path from point $P_3$ to point $P_2$ in Fig. 3(b) ($P_i \in$ Points of Interest POI), representing the experiment Jaeger et. al. [3]. As in analogy to (6), the kernels of two entangled photons from their common origin, e.g., point in a nonlinear crystal, to each arbitrary point of their trajectories can be described by multiplying their kernels, because each path of the entangled partners fulfills the criteria common origin and propagation by "principle of least action". So, the virtual composite path between entangled quanta obeys the "principle of least action", too. The impartible system of entangled quanta fulfills the "principle of least action" between all combinations of arbitrary points of their trajectories.

The source of entangled quanta is represented by the grey square in the middle in Fig. 3(b). (Mapping between a MZI and Jaeger et. al.: $BS_1 \leftrightarrow H_1$; $E_1 \leftrightarrow L_1$; $E_6 \leftrightarrow U_2$; $\Phi \leftrightarrow \Phi_2$; additional $\Phi_1$ in Jaeger et. al.; …).

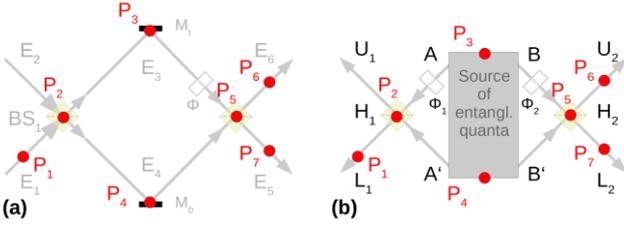

*Fig. 3: These two sketches give an idea, how the spatial superposition of a MZI in (a) is related to the experiment of entangled photons of Jaeger et. al. in (b). The amplitude of second-order interference of entangled photons in equation (20) includes ($\Phi_1 + \Phi_2$) in the absolute value of the phase difference.*

As seen in the introduction, the dimension of the path integral is a phase. Therefore, such a phase is a time independent factor and can also represent arbitrary way lengths, which is essential to describe delayed choice arrangements or ghost imaging. The kernel $K(P_6,K_1)_{top}$ can be split into two kernels, $K(P_1,P_3)_{top}$ and $K(P_6,P_3)_{top}$,

$$K(P_1, P_3)_{top} = \int_{P_3}^{P_1} exp\left[\frac{i}{\hbar}S[P_1,P_2] + \frac{i*0_{H_1}}{2} + i*\phi_1 + \frac{i}{\hbar}S[P_2,P_3]\right] Ðx(t) , \quad (16)$$

$$K(P_6, P_3)_{top} = \int_{P_3}^{P_6} exp\left[\frac{i}{\hbar}S[P_6,P_5] + \frac{i*\pi_{H_2}}{2} + i*\phi_2 + \frac{i}{\hbar}S[P_5,P_3]\right] Ðx(t) , \quad (17)$$

and kernel $K(P_6,K_1)_{bottom}$ into two kernels, $K(P_1,P_4)$ and $K(P_6,P_4)$, respectively.

$$K(P_1, P_4)_{bottom} = \int_{P_4}^{P_1} exp\left[\frac{i}{\hbar}S[P_1,P_2] + \frac{i*\pi_{H_1}}{2} + \frac{i}{\hbar}S[P_2,P_4]\right] Ðx(t) . \quad (18)$$

$$K(P_6, P_4)_{bottom} = \int_{P_4}^{P_6} exp\left[\frac{i}{\hbar}S[P_6,P_5] + \frac{i*0_{H_2}}{2} + \frac{i}{\hbar}S[P_5,P_4]\right] Ðx(t) . \quad (19)$$

As the phase of a path integral has no information of a time order any more, it is possible to describe the phase solution space of the entangled system, which fulfills the principle of least action for the entangled system, by the probability amplitudes of the combination of the probability amplitudes of the entangled partners. The probability amplitude of second-order interference $A(P_1,P_6)$ is now

$$A(P_1, P_6) = \frac{1}{\sqrt{2}}(K(P_1,P_3)_{top}K(P_6,P_3)_{top} + K(P_1,P_4)_{bottom}K(P_6,P_4)_{bottom}) , \quad (20)$$

**including $\Phi_1$ and $\Phi_2$** in the corresponding kernels $K(P_1,P_3)_{top}$ and $K(P_6,P_3)_{top}$, respectively.

By using the notation of the experiment of Jaeger et. al., the whole set of "second-order interference probabilities" consists of four equations,

$P(U_1, U_2) = (A(U_1, U_2))^2 = ¼ [1 − \cos(\Phi_1 + \Phi_2)]$
$P(U_1, L_2) = (A(U_1, L_2))^2 = ¼ [1 + \cos(\Phi_1 + \Phi_2)]$
$P(L_1, U_2) = (A(L_1, U_2))^2 = ¼ [1 + \cos(\Phi_1 + \Phi_2)]$
$P(L_1, L_2) = (A(L_1, L_2))^2 = ¼ [1 − \cos(\Phi_1 + \Phi_2)]$, (21)

which define the shared phase solution space of second-order interference, considering the **nonlocal impact** of phase shifter $\Phi_1$ **and** phase shifter $\Phi_2$. So, the shared phase solution space of entangled quanta is spanned by nonlocal phases. (comprehensible also in [4], [5], especially in delayed choice experiments like [6], [7])

The formalism to calculate the non-coincidence single detector probabilities by summing over the corresponding counts of the detectors of the entangled subsystem is misleading and eliminates the interference terms. Eliminating the interference terms (e.g., trace of reduced density matrix, summing over probabilities) wipes off the existing "nonlocal information".

$$P(L2) = P(U1,L2) + P(L1,L2) = \\ = ¼ [1 + \cos(\Phi_1 + \Phi_2)] + ¼ [1 − \cos(\Phi_1 + \Phi_2)] = \\ = 1/2 . \quad (22)$$

Among other entangled properties of such systems, the shared (common) phase solution space is the most important one. To allow the measurement of the shared phase solution space in just one place, it will be key to quantum information technology to eliminate the necessity of coincidence measurement.

A very impressive example of such masking of second-order interference to look like first-order interference and, therefore, to measure second-order interference in just one spacetime point, is found in Lemos et. al. [5]. This experiment shows the importance of the shared phase solution space of entangled quanta.

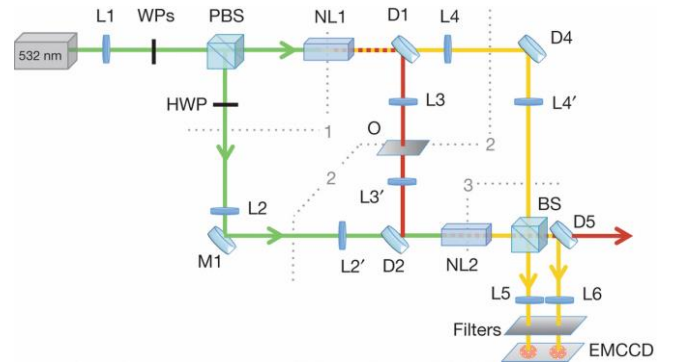

*Fig. 4: Shows the consequence of second order interference, the silicon plate with etched and non-etched regions (a cat) is inserted as object O in path D1-D2. The optical path length difference between these regions is half of the wavelength of the idler photons.*
*© GB Lemos et al. Nature 512, 409-412 (2014) doi:10.1038/nature13586*

In Figure 3 in Lemos et. al. the cardboard cut-off "Schrödingers cat" appears in an "... *interesting single-particle superposition, ...*" [5], which is in fact a consequence of the shared phase solution space and the which-way information. Checking the phases of the entangled idler photons (see Fig. 4 to see Figure 2 in Lemos et. al.), one can see, that idlers originated in NL1 and idlers originated in NL 2 have the same phase, so the phases of the idler photons act as a global phase (see Heuer et. al. [8] for the phase context before the nonlinear crystals in this experiment). In this case, both idler paths contribute the same phase (NL1-idler and NL2-idler) to the combined path integrals. Therefore, the interference in BS2 appears as first-order interference, despite it is still a second-order interference. The proof of the existence of second-order interference was made by placing a silicon "Schrödingers cat" ("... *the difference in optical path length for the etched and non-etched regions corresponds to a relative phase shift of π, ...*")[5] in one of the idler modes. (see Fig. 5 to see Figure 4 in Lemos et. al.)

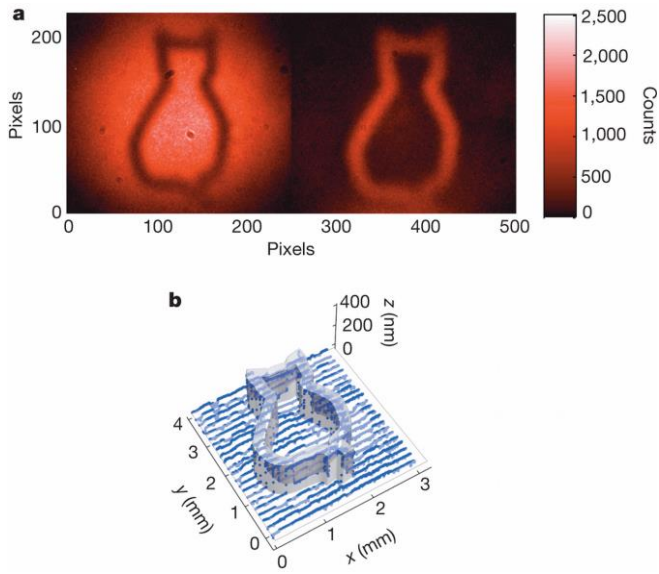

*Fig. 5: Phase image of an object opaque to 810-nm light in Lemos et. al.[5]. The etched and non-etched regions of the object are shown. The contour of the cat causes an additional phase shift in one of the idler modes, which changes the phase difference measured after BS by π.*
© GB Lemos et al. Nature 512, 409-412 (2014) doi:10.1038/nature13586

Now the different phases of the idler modes become visible, because the "first-order interference" intensity, which is in fact still second-order interference intensity, flips between the modes g and h (see Lemos et. al. for details). This is a proof, that second-order interference always depends on the phase solution space of the complete entangled quantum system, including paths of an existing spatial superposition, not just the phase of one quantum alone as it looks like sometimes. [9]

### IV) Complementarity

On the way to benefit from the non-locality of entangled systems in information technologies, the distinction between first-order interference and second-order interference has to be precise.

In the context of complementarity, an entity, which can only be understood by contradictory properties, is nevertheless the same entity in each observation. Only the measurable properties can be complementary. If the entities are different, then complementarity is not the optimal concept. To change from first-order interference to second-order interference someone has to entangle at least two "first-order capable" entities to an entangled system.

First-order interference is measurable only for independent single quanta. The entity is a single independent quantum. The amplitude in (13) describes the transition of only one photon. Second-order interference is measurable only for entangled systems of quanta. A system of entangled photons consists of more than one single photon, which is a different type of an entity. The amplitude in (20) describes the transition of two photons. There is no order for the multiplication of the phases based on a time order of the corresponding kernels within this amplitude, because the resulting phases of the path integrals have not time dimension anymore.

So, first-order and second-order interference are based on different types of entities, which is in contrast to Bohrs understanding of complementarity. [10]

Why is this specification necessary? By mixing the types of interference in the literature like "*Second-order interference is observed in the superposition of signal photon ...* " [9], in the discussion of e.g., delayed choice experiments the nonlocal effect of entanglement can get lost.

### V) Conclusion and Outlook

The path integral analysis shows, that a "Complementarity of one-particle and two-particle interference" described in Jaeger et. al. [3] does not exist (see also "The phenomenon then becomes a second-order effect." in [9]). Assuming the existence of a spatial superposition, unentangled quanta show first-order interference, entangled quanta show always second-order interference. Depending on the setup of the experiment, second-order interference can appear to be first-order interference, but it isn't.

The rule for a single quantum by Feynman:

"*Amplitudes for events occurring in succession in time multiply.*"

has to be extended to meet entangled quanta and their shared phase solution space:

"*Amplitudes for events of a single quantum of an entangled system of quanta occurring in succession in time multiply. Resulting amplitudes (resulting actions S) of single quanta of entangled systems of quanta multiply with their entangled partners amplitudes without considering the succession in time*".

The shared phase solution space of entangled quanta, representing also the foundation of second-order interference, is spanned by nonlocal phases, which is essential to preserve the principle of least action within entanglement.

Assume that beam splitter $H_1$ in Jaeger et. al. [3] is at the distance of 1m to the source of entangled photons and $H_2$ is one light year away. The phase shifters $\boldsymbol{\Phi_1}$ and $\boldsymbol{\Phi_2}$ are close to the corresponding beam splitters. As the equations in (21) show, both phase shifters contribute to the relative phase difference of second-order interference, determining the probability of which out port of the beam splitter $H_1$ has to be taken. As this is a consequence of the principle of least action and the probability has to be the same in all reference frames, the quantum correlation depends on deliberate settings of spacelike phase shifters. Even in a superposition of time orders, depending on a deliberate setting is more likely a causal effect than a simple statistical correlation.

This, and the special cases of such quantum causal relations in time like settings will be described in the next step in a further publication.

And, the description of the nonlocal experiment, which offers interference measurement of second-order interference without coincidence protocols, will follow after this step, being the last step of demonstrating nonlocal quantum causal relations.

Through simultaneous fulfillment of the self-consistency principle of causation by the guarantor "principle of least action", the utilization of the nonlocal phases of the shared phase solution space of entangled quanta opens the door to new quantum communication protocols and quantum information processing.

**Acknowledgment**

Special thanks to A. H. for requesting a written basis for discussion. Moreover, I am still very, very grateful to Helmut Rauch for the special encouragement "stay on course" as long as there are no contradictions. And most of all, I want to thank Siegfried Fussy for all these valuable and critical discussions.

This work was partly supported by Intellectual Capital and Asset Management GmbH, Austria.